\documentclass{epl}
\usepackage[T1]{fontenc}
\usepackage{graphics}

\title{Interplay of frequency-synchronization with noise: current resonances,
giant diffusion and   diffusion-crests}
\shorttitle{Interplay of frequency-synchronization with   noise}

\author{D. Reguera\inst{1}$^{*}$ \and  P. Reimann\inst{2} \and  P. Hänggi\inst{2} \and  
J.M. Rubí\inst{1}}
\shortauthor{D. Reguera \etal}
\institute{
  \inst{1} Departament de F\'{\i}sica Fonamental, Facultat de F\'{\i}sica,
Universitat de Barcelona,\\
Diagonal 647, 08028 Barcelona, Spain.\\
$^{*}$e-mail: davidr@precario.ffn.ub.es \\
  \inst{2} Universität Augsburg, Institut für Physik, Universitätsstrasse 1, D-86135 Augsburg,
  Germany}
  
\pacs{05.40.-a}{Fluctuation phenomena, random processes, noise, and   Brownian motion}
\pacs{05.45.Xt}{Synchronization; coupled oscillators}
\pacs{05.60.-k}{Transport processes}

\begin{document}

\maketitle
\begin{abstract}
We elucidate how the presence of noise may significantly interact with the synchronization
mechanism of systems exhibiting frequency-locking. The response of these systems
exhibits a rich variety of behaviors, such as resonances and   anti-resonances
which can be controlled by the intensity of noise. The transition between different locked regimes
provokes the development of a multiple enhancement of the effective diffusion. This diffusion 
behavior is accompanied
by a crest-like peak-splitting cascade when the distribution of the lockings is self-similar,
as it occurs in periodic systems that are able to exhibit a Devil's staircase sequence of
frequency-lockings.
\end{abstract}
The phenomenon of frequency synchronization or frequency-locking is generic for
nonlinear  dynamical systems
where two or more frequencies compete. It has been observed in a great variety
of situations including the cases of the driven pendulum, charge-density-waves
\cite{sherwin}, chemical reactions \cite{swinney} or Josephson junctions \cite{shapiro},
to mention just a few. Its main characteristics is the appearance of a complex
distribution of plateaus or staircase structure in the response of these systems.
The presence of the plateaus is the signature of the locking at different frequencies
which are rational multiples of the driving frequency.

A simple model exhibiting frequency-locking is the periodically driven washboard
potential. This type of  potential is particularly interesting since its ubiquity
and  simplicity makes it the archetype for modeling transport in periodic systems.
A great variety of condensed-matter systems can be modeled using this potential.
We may refer the cases of Josephson junctions \cite{Josephson}, charge-density-waves
\cite{CDW}, superionic conductors \cite{superionic}, rotation of dipoles in
external fields \cite{dipolos}, phase-locked loops (PLL) \cite{PLL}, diffusion
on surfaces \cite{diff_on_surf} or separation of particles by electrophoresis
\cite{electroph}, to mention just a few \cite{risken}. When the washboard
potential is periodically driven it exhibits a great variety of non-linear phenomena
including phase-locking, hysteresis \cite{hysteresis}, stochastic resonance
\cite{SR} and   chaos \cite{chaos} .

Our purpose in this Letter is to report interesting new phenomena which take place in
periodically driven multistable systems arising as a consequence of the cooperative effect
between synchronization and   noise. We shall demonstrate that the interplay between the 
frequency-locking
and  the noise gives rise to a multi-enhancement of the effective diffusion,
and  a rich behavior of the current as well, including partial suppression and   characteristic resonances.

For the sake of simplicity, we will start our analysis considering the model
of overdamped Brownian motion of a particle in a washboard potential with a modulated tilt.
 In scaled units, the dynamic equation is given by

\begin{equation}
\label{modelo}
\frac{dx}{dt}=-\sin (x)+F+A\cos (\omega t)+\xi ,
\end{equation}
where \( F \) is the tilt, \( A \) is the amplitude of the periodic input,
and  \( \omega  \) its angular frequency. The system is under the influence of a Gaussian
white noise \( \xi  \) of zero mean and   delta-correlated \( \left\langle \xi (t)\xi (t')\right\rangle =2D\delta (t-t') \),
with \( D \) defining the noise level. Transport characteristics can be described
in terms of two main quantities: the average velocity, or current, defined in
the long time limit as

\begin{equation}
\label{velocidad}
\left\langle \frac{dx}{dt}\right\rangle \equiv \left\langle v \right\rangle\equiv \begin{array}{c}
lim\\
t\rightarrow \infty 
\end{array}\frac{\left\langle x(t)\right\rangle}{t},
\end{equation}
 which is independent on the initial condition \( x(0) \). The effective
diffusion coefficient  is defined as 

\begin{equation}
\label{coeficiente de dif efectivo}
D_{eff}\equiv \begin{array}{c}
lim\\
t\rightarrow \infty 
\end{array}\frac{1}{2t}\left\langle \left[ x(t)-\left\langle x(t)\right\rangle \right] ^{2}\right\rangle .
\end{equation}

\begin{figure}
{
\resizebox*{0.47\columnwidth}{!}{\includegraphics{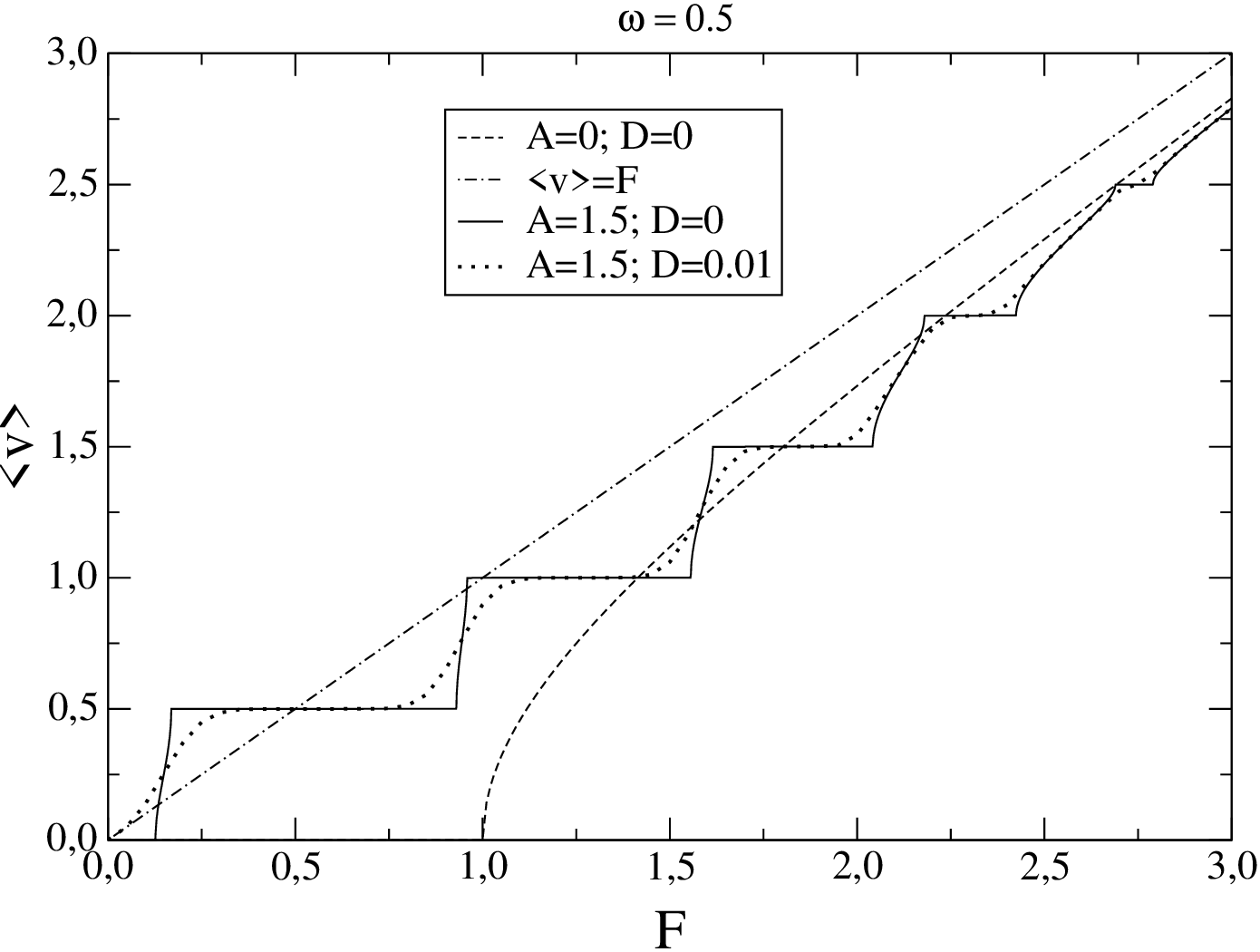}}
\hfill\resizebox*{0.47\columnwidth}{!}{\includegraphics{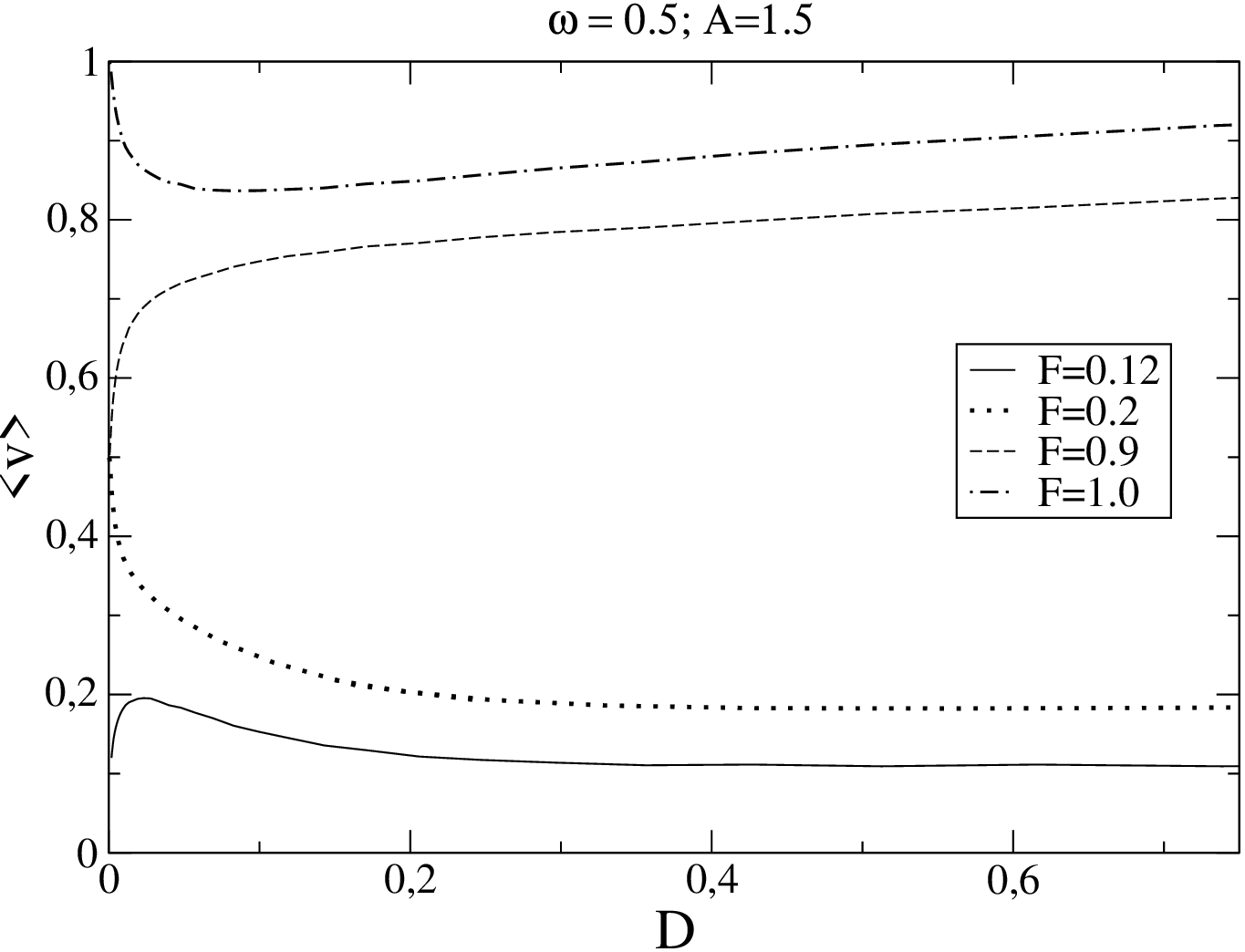}} }
\caption{Average velocity as a function of the tilt \protect\( F\protect \),
for angular driving frequency \protect\( \omega =0.5\protect \) for different driving
strengths and   noise intensities. The deterministic behavior in absence of driving
 is depicted with the dashed line; the Shapiro step behavior is shown by the solid line. 
 Small noise smooths the Shapiro steps (dotted line).\label{shapiro}}
\caption{Velocity as a function of the level of noise for a driving strength 
\protect\( A=1.5\protect \) and   angular frequency 
\protect\( \omega =0.5\protect \) for different values of the tilt \protect\( F\protect \).\label{v(D)}}
\end{figure}

A main feature of the deterministic dynamics of the model is the appearance
of a multiplicity of plateaus in the current when represented against the tilt,
which have been referred to as Shapiro steps \cite{shapiro} (see Fig. \ref{shapiro}).
This trait is a consequence of the locking of the particle velocity at
the harmonics of the driving frequency. Naively speaking, the
frequency-locking occurs because the particle tends to synchronize its motion with
the periodic force to overcome an integer number of wells during one cycle of
the force. The size of the jumps in the current is  \( \Delta \left\langle v
\right\rangle=\omega  \). The deterministic dynamics of this model has been
widely characterized in the literature \cite{kautz}, and   the existence of mode-locking
has been corroborated experimentally \cite{sherwin},\cite{shapiro}.

A remarkable consequence of the locking is that the response of the system can
be controlled by means of the modulated component of the tilt (for instance,
mode-locked Josephson junctions are used as a voltage standard \cite{kautz_chaos}).
Particularly interesting is the fact that, for some values of the parameters,
one can achieve values of the velocity larger than the tilt \( F \), as occurs
for \( F=0.25 \), and   \( A=1.5 \) in the example depicted in Fig. \ref{shapiro}.

These peculiar nonlinear characteristics of the deterministic dynamics gives rise to interesting
transport phenomena when noise is present. There has been a pronounced interest in the
effect of noise in this model \cite{stratonovich}, especially in the context of Josephson junctions,
laser-gyroscopes and   charge-density-waves. The linear and   nonlinear response
has been characterized as a function of the tilt \cite{stratonovich}, the driving frequency or the
amplitude \cite{coffey}-\cite{jung}, and  such intriguing phenomena as the reduction
of the low-frequency broadband  noise level in the locked regime of charge-density-waves
have been measured\cite{sherwin} and   described \cite{wiesenfeld}. Here we will
focus on two different, previously unexplored aspects: the response as a
function of the noise level and   the behavior of the effective diffusion in
terms of the control parameter (the tilt or the amplitude). Our results have
been obtained from numerical simulations of Eq. (\ref{modelo}).

When the velocity is analyzed in terms of the noise level, it exhibits different
regimes, which can be tuned through the value of the tilt \( F \), as depicted with  Fig.
\ref{v(D)}. One of these regimes, corresponding to the curve \( F=0.9 \)
in Fig. \ref{v(D)}, reflects a monotonic increase of the current with the
intensity of noise (see also below). This behavior just mimics the situation which occurs 
in the absence of driving (not shown). Then, the addition of noise facilitates the escape of the particle
trapped in the potential, thus increasing its velocity downhill. However, when
the periodic driving is present and   synchronization occurs, a more rich phenomenology
emerge. For instance, the opposite behavior can also be found, i.e., a monotonic
decrease of the current {\it vs.} increasing  noise intensity (as for \( F=0.2 \) in Fig.
\ref{v(D)}). Moreover, for another range of values of the tilt, the behavior
is no longer monotonic: the current exhibits either an anti-resonance or a resonance, evidenced
through the presence of  a minimum (as for \( F=1.0 \)
in Fig. \ref{v(D)}) or a maximum (\( F=0.12 \) in Fig. \ref{v(D)}), respectively, at characteristic
values of the noise level.

The jumps in the current, which are the result of the locking of the velocity at integer
multiples of the driving frequency, yield the underlying mechanism for the appearance
of the different regimes. Their existence can be qualitatively understood from
the analysis of the deterministic dynamics (see Fig. \ref{shapiro}). A small
amount of noise tends to smooth the steps because the noise helps the particle
to escape from its trapped, phase-locked state. Consequently, at small values
of the noise, the current increases for values of the tilt $F$ on the right half
 of the plateaus
(i.e., before the jump for the velocity has occurred)  and   decreases
on the other half of the plateaus (i.e., after the jump has occurred). For high noise levels,
 the particle does not notice the presence
of the potential and   its velocity thus tends to a limiting value which correspond
to the value of the tilt \( F \), i.e. $\left\langle v \right\rangle\rightarrow F$ as $D \rightarrow \infty$.
 These two limiting tendencies determine the
appearance of the different regimes. For instance, a monotonic increase of the current may
occur for values of the tilt $F$ on  the right half
of  synchronization plateaus (as for \( F=0.9 \) in Fig. \ref{v(D)}) because
  the addition of noise consistently increases the velocity up to the
 limiting value of $\left\langle v \right\rangle=F$. Similarly, the anti-resonance occurs for values of the tilt on the
 left half of the plateaus (as for \( F=1.0 \) in Fig. \ref{v(D)}), because the smoothing
  of the steps leads to an initial decrease of
 the current with the noise, followed by an increase in order 
 to reach the asymptotic  $\left\langle v \right\rangle\rightarrow F$. The occurrence of the remaining regimes requires special
 conditions. In particular, a maximum of the velocity or a monotonic decreasing regime may
  only appear for values of the tilt 
  before and   after a step in which $\left\langle v \right\rangle$ exceeds the value $\left\langle v \right\rangle=F$ for $D=0$ (as \( F=0.12 \) and   \( F=0.2 \) in
  Fig. \ref{v(D)}, respectively). The necessity of this requirement can be inferred
   from the above reasoning.
  It is also important to 
remark that the behavior of the
current {\it vs.} the tilt $F$ is similar for values of the driving strength $A<1$,
 thus manifesting that
these anomalous behaviors are not a peculiarity of strong driving. 

\begin{figure}[t]
{ \resizebox*{0.47\columnwidth}{!}{\includegraphics{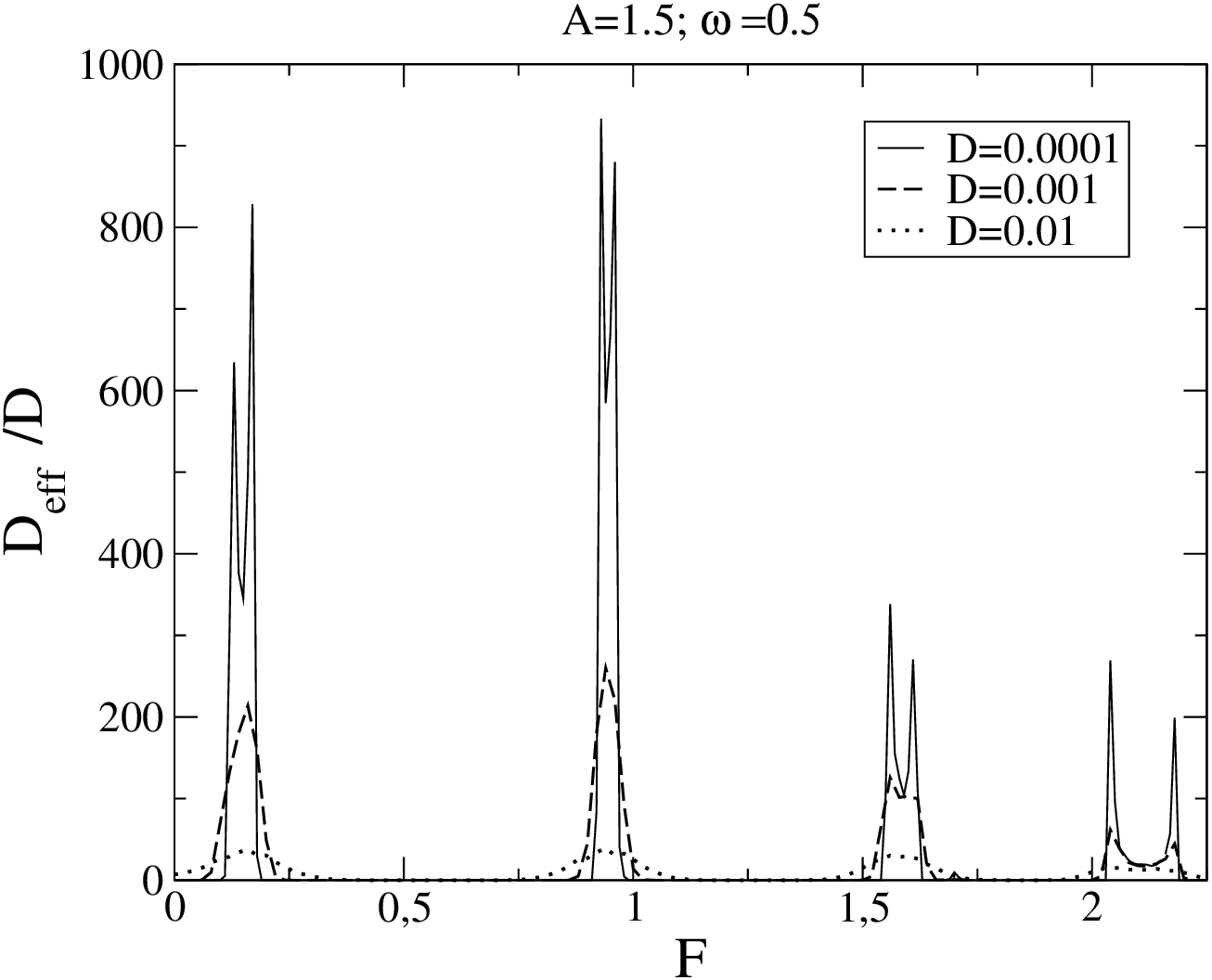}}
\hfill\resizebox*{0.47\columnwidth}{!}{\includegraphics{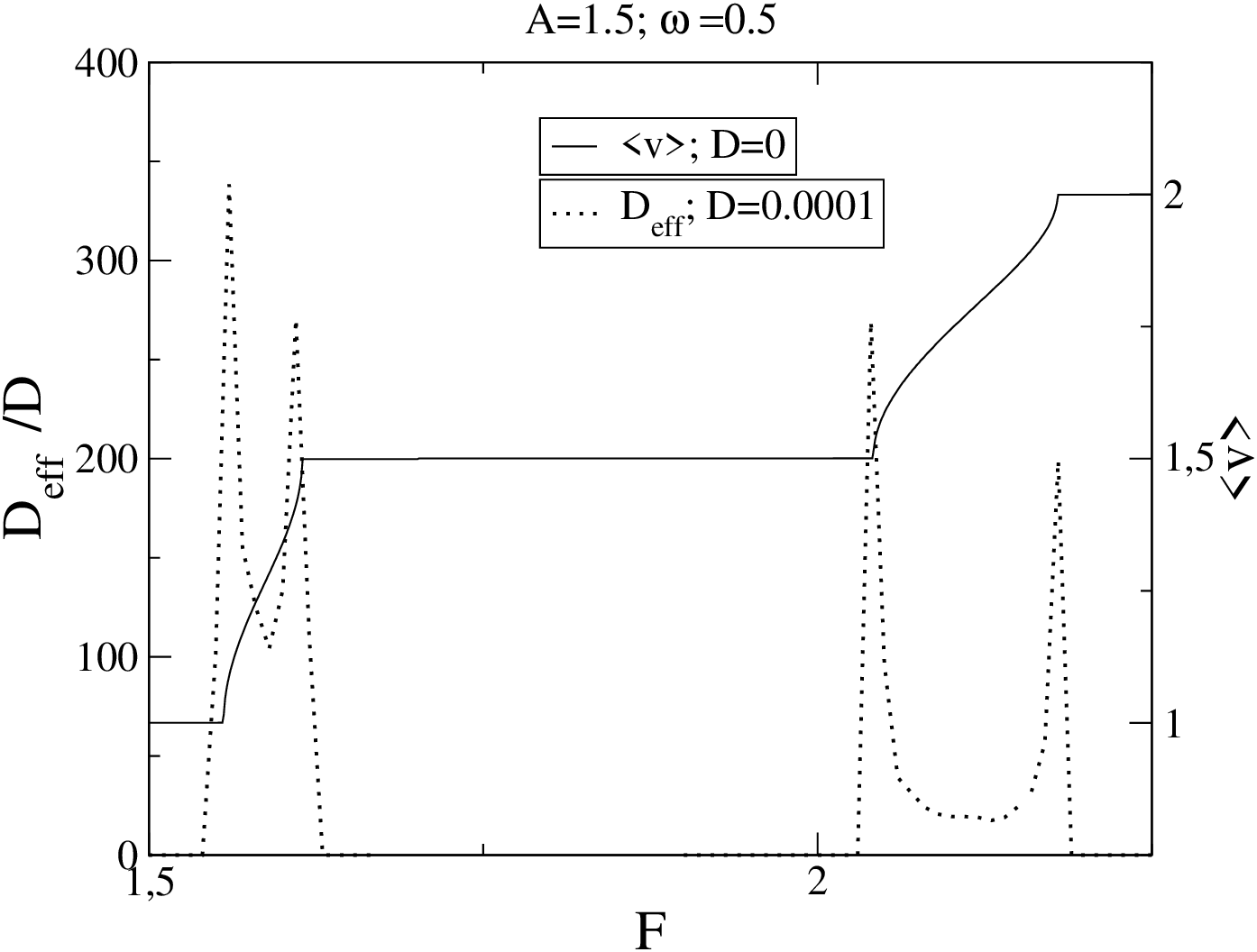}}}

\caption{Effective diffusion \protect\( D_{eff}/D\protect \) as a function of the tilt
\protect\( F\protect \), for a driving strength \protect\( A=1.5\protect \), angular 
frequency \protect\( \omega =0.5\protect \)
for  different values of the level of noise \protect\( D\protect \).\label{deff}}
\caption{Close-up of the behavior of \protect\( D_{eff}/D\protect \) in Fig. \protect{\ref{deff}}
superimposed to the plot \protect\( \left\langle v \right\rangle\protect \) \protect{\it vs}.
\protect\( F\protect \). The peaks in \protect\( D_{eff}\protect \) occur at the onset 
of transition between
the locked and   non-locked regimes.\label{splitting}}
\end{figure}
 
 The transition between the locked and   non-locked 
regimes in the deterministic dynamics gives rise to a peculiar behavior of
the effective diffusion. In the numerical simulations of Eq. (\ref{modelo})
one observes the presence of multiple peaks in the effective diffusion when
represented as a function of the tilt, as depicted in Fig. \ref{deff}. As the
intensity of noise diminishes, the ratio \( D_{eff}/D \) significantly increases,
which is the signature of an enhancement of the effective diffusion. The location
of the peaks corresponds to the values of the tilt \( F \) at which the jumps in the
current {\it vs.} \( F \) occur. Consequently, by means of the periodic forcing one can control
the conditions under which the enhancement of diffusion occurs. That is, by
selecting the proper amplitude or frequency of the periodic driving, one can
obtain a significant enhancement of the diffusion at an arbitrary value of the
tilt. Moreover, the effect is {}``robust{}'' in the sense that there exist
a multiplicity of values of the tilt for which an enhancement of the diffusion
can be obtained. 

The mechanism for the enhancement of the effective diffusion can be traced back to
the extreme sensitivity of the dynamics upon the addition of a small amount
of noise.  In the frequency-locked regime, the motion of the particle is enslaved by the
periodic forcing, hence  the effective dispersion is small.
 In fact, 
the random perturbations are effectively suppressed in the locked regime 
\cite{sherwin},\cite{wiesenfeld}, 
and  the effective, barrier-activated diffusion $D_{eff}$ is smaller
than the free diffusion $D$. On the contrary, in the non-locked regime, the behavior of the
stochastic particle dynamics becomes strongly diffusive with large fluctuations.
These extreme fluctuations emerge because of a drift-assisted splitting mechanism of the
Brownian particle dynamics. The giant enhancement of the effective diffusion 
takes place close to the parameter region where the particle gets rid of the captivity of the periodic driving. 
 It is precisely at those transition points where the dynamics of the particle becomes particularly sensitive to the addition
  of a small dose
 of noise. This happens because the synchronization is efficiently nourishing this drift-assisted splitting mechanism.
 Similar features been reported in Refs. \cite{reimann},\cite{marchesoni} for
the tilted washboard potential in the absence of periodic modulation, and   in
Refs. \cite{gang},\cite{schreier} for other systems at slow driving.

Qualitatively, it is quite suggestive that a very sensitive dependence of 
    the current $\langle v\rangle$ upon small changes of the static force $F$
    will be intimately connected with a very sensitive dependence of the
    particle trajectory $x(t)$ on the random force $\xi (t)$, resulting in
    a greatly enhanced diffusion $D_{eff}/D$. In other words, we may expect
    a qualitative (but not quantitative) proportionality between
    $d\langle v\rangle/dF$ and $D_{eff}/D$, which is indeed confirmed by
    our numerical findings. In particular, a careful analysis of the effective diffusion at very low levels of noise reveals
the existence of a splitting of the peaks, showing that the enhancement occurs
at \emph{every} transition between locked and   non-locked regimes, and   {\it
vice versa}.
This characteristic is illustrated in Fig. \ref{splitting}. In the limit $D \rightarrow 0$ 
 there is no further splitting, and   the ratio $D_{eff}/D$ assumes giant values while
 at the same time the corresponding peak width
narrows, thus 
tending to a collection of very sharp singularities  which are located at every onset and  every end of the locked
regimes.  

This effect is not restricted to the behavior of the effective diffusion as a function of the
tilt \( F \). The enhancement of the effective diffusion also takes place at the onset
of frequency-locking as a function of  other control parameters, such as the amplitude 
 \( A \)
and  the angular frequency $\omega$. 

\begin{figure}
{\resizebox*{0.47\columnwidth}{!}{\includegraphics{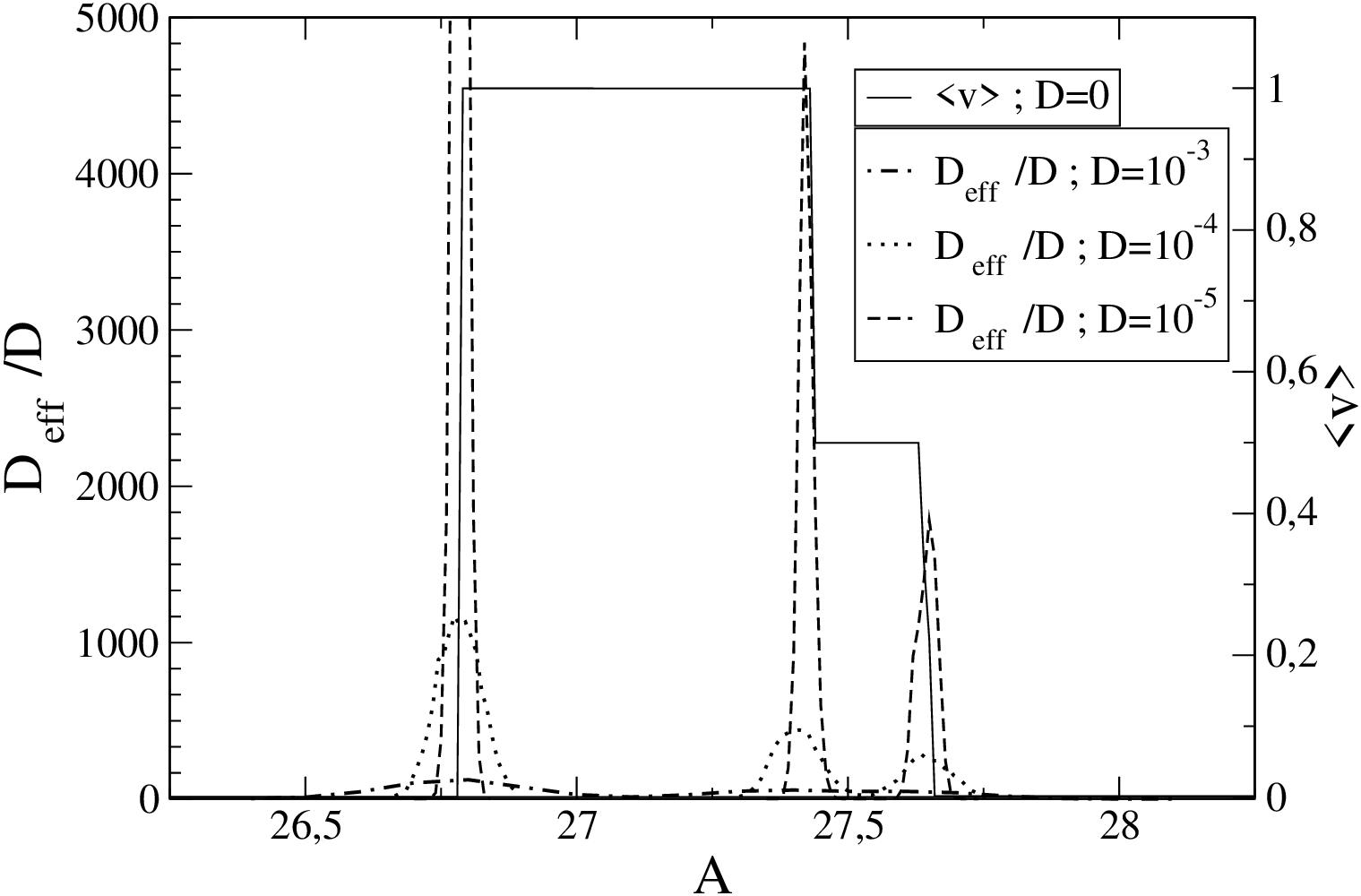}} 
\hfill\resizebox*{0.47\columnwidth}{!}{\includegraphics{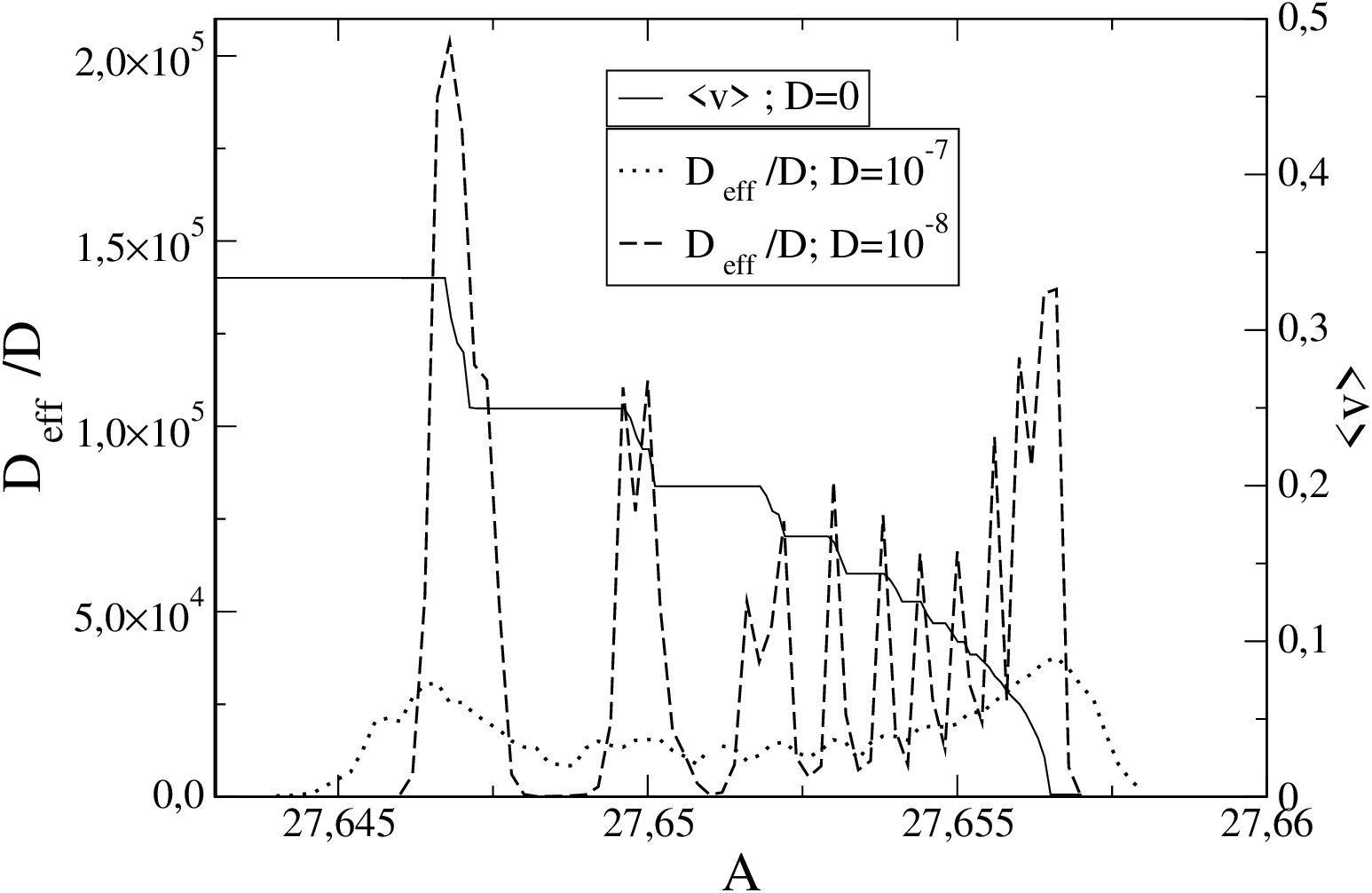}}}

\caption{Effective diffusion \protect\( D_{eff}/D\protect \) as a function of the amplitude
strength \protect\( A \protect \), for the model described by Eq. (\ref{prost})
at different values of the level of noise \protect\( D\protect \). The plot
of the deterministic response \protect\( \left\langle v \right\rangle\protect \) {\it vs.} \protect\( A \protect \)
is overlaid.\label{devil1}}
\caption{Magnified view of parts of Fig. \protect{\ref{devil1}}. The plot 
\protect\( \left\langle v \right\rangle\protect \) {\it vs.}
\protect\( A \protect \) exhibits a Devil's staircase structure. The
splitting of the peaks of the effective diffusion, occurring at the onset of
transition between different locked regimes, becomes increasingly more manifested as we decrease
the level of noise.\label{devil2}}
\end{figure}

The results reported above reveal the basic mechanism for the occurrence
of multi-enhan\-ce\-ment of the effective diffusion. In the model we have implemented,
the locking  takes place only at integer values of the driving frequency. We
can now analyze the implications of our results in more complex models exhibiting
different types of synchronization mechanisms. A particularly interesting situation
is the case of potentials giving rise to a self-similar distribution of steps,
known as Devil's staircase \cite{bak}. In these systems, in the absence of noise the phase locking
occurs at every rational value of the driving frequency \( \left\langle v \right\rangle=\frac{n}{m}\omega  \),
with \( n, \) \( m \) being integers. In that case, one has a self-similar
distribution of transitions between locked and   non-locked states, suggesting the occurrence of a 
corresponding cascade of splitting of the peaks in the effective diffusion in
the presence of a small amount of noise. In Figs. \ref{devil1} and   \ref{devil2}, we have calculated
the effective diffusion by solving numerically the model suggested in Refs.
 \cite{adjari},\cite{bartussek}

\begin{equation}
\label{prost}
\frac{dx}{dt}=-\varepsilon \frac{dU}{dx}+ A \cos (2\pi t)+\xi ,
\end{equation}
where the amplitude is \( \varepsilon =5 \), the periodic potential is
 \( U(x)=\left[ \cos (2\pi x)-0.5\sin (4\pi x)\right] /2\pi  \), 
 and   \( \xi  \) is a Gaussian
white noise with zero mean and   \( \left\langle \xi (t)\xi (t')\right\rangle =2D\delta (t-t') \).
For these values of the parameters, it has been shown in Ref. \cite{adjari} that the
model exhibits a Devil's staircase structure
in the current behavior, see the \( \left\langle v \right\rangle \) {\it vs.} \( A  \) plots of Figs. \ref{devil1} and   \ref{devil2}.
Those figures clearly confirm the expected  progressive  splitting of peaks of the effective diffusion
which occurs as the  noise intensity is reduced. It reveals the rich structure
that the interplay between synchronization and   noise generates in this case.
The superimposed plots of \( \left\langle v \right\rangle \) {\it vs.} \(A  \) corroborate that the
enhancement of the effective diffusion occurs at \emph{every} onset of transition
between different locked regimes. The overall behavior for  $D_{eff}$ thus
assumes a hillcrest-like form, {\it cf.} Fig. \ref{devil2}. This novel characteristic bearing
of $D_{eff}$ {\it vs.} driving strength $A$ shall be termed a {\it
diffusion-crest}. We remark that in the rather different context of deterministic 
            dynamics in discrete time (chaotic maps) a diffusion coefficient
            with a Devil's staircase-like behavior has been revealed in 
            \cite{kla95}, while in our case the diffusion coefficient exhibits
            an even more spectacular behavior, approaching qualitatively
            the {\em derivative} of a Devil's staircase for asymptotically 
            weak noise.

In conclusion, we have shown that the cooperation between noise and   
frequency-locking inherent to the deterministic dynamics of periodically driven
systems gives rise to the appearance of a rich transport phenomenology. When the response
of a systems assumes a staircase structure, the addition of noise leads to
the occurrence of counterintuitive  phenomena. One then observes the existence
of regimes in which the current is suppressed or the presence of resonances
or anti-resonances in terms of the noise level. The lowering of the noise level
reveals the existence of a very peculiar behavior of the effective diffusion
including a strongly multi-enhancement and  a crest-like splitting of the peaks which can
become self-similar in the case of a locking behavior of the Devil's staircase type.
 This behavior evidences
the increase of the sensitivity of the system to small perturbations at the
transition between different locked and  non-locked regimes. The  multiple diffusion 
enhancement
is then a general phenomenon occurring in systems for which frequency-synchronization
occurs and  the response to an input assumes a step-like structure. This new
phenomenology implies interesting consequences in transport theory of nonlinear
systems and  the findings can in principle be applied to actual situations of practical
interests. The existence of diffusion-crests provides a promising mechanism for the
selective control of diffusion. This mechanism could be useful, for example, to 
improve and control selectively the release of drugs in biological tissues
\cite{drug-release}. For
field-responsive systems \cite{rosenweig},\cite{fieldr}, {\it i.e.}  electro-
and magneto-rheological fluids,  the giant and controlled enhancement of the diffusion
would give rise to a huge increase of the rotational viscosity with potential applications
to magnetic dampers.
\acknowledgments
This work was supported by the exchange program of the Deutscher Akademischer Austauschdienst
(P.H, P.R.) and  the Acciones Integradas Hispano-Alemanas under HA1999-0081
(D.R., J.M.R.),
the Generalitat de Catalunya (D.R.), DFG-Sachbeihilfe HA1517/13-4 and 
           the Graduiertenkolleg GRK283 (P.H., P.R.).


\end{document}